\pdfoutput=1
\documentclass[]{aa}
\usepackage{graphicx}
\usepackage{latexsym}
\usepackage{graphicx}
\usepackage{subfigure}
\usepackage{longtable}
\usepackage{supertabular}
\usepackage{natbib}
\usepackage[latin1]{inputenc}
\usepackage[thinspace]{SIunits}
\usepackage{graphicx}
\usepackage{graphics}
\usepackage{epsfig}

\newcommand{\bea}{\begin{eqnarray*}}          
\newcommand{\eea}{\end{eqnarray*}}

\begin{document}

\title{Plane--mirroring anomaly in the cosmic microwave background maps}
\author{V.G.Gurzadyan\inst{1},  T.Ghahramanyan \inst{1},  A.L.Kashin\inst{1}, \\
H.G.Khachatryan\inst{1}, A.A.Kocharyan\inst{1,2}, H.Kuloghlian\inst{1}, D.Vetrugno\inst{3} and G.Yegorian\inst{1}}

\institute
{
\inst{1} Yerevan Physics Institute and Yerevan State University, Yerevan, Armenia\\
\inst{2} School of Mathematical Sciences, Monash University, Clayton, Australia\\
\inst{3} University of Lecce, Lecce, Italy
}

\date{Received (\today)}

\titlerunning{Large scale plane-mirroring in CMB maps}

\authorrunning{V.G.Gurzadyan et al.}

\abstract{
The plane-mirror symmetry previously noticed in the Cosmic Microwave Background (CMB) 
temperature anisotropy maps of Wilkinson Microwave Anisotropy Probe is shown to possess
certain anomalous properties. The degree of the randomness determined by the Kolmogorov stochasticity parameter
in the both symmetry regions appears to have identical values which, however, essentially differ from the corresponding
values for other sky regions. If the mirroring were of cosmological origin, this would imply either additional
randomizing properties in those directions of the Universe or their different line-of-sight depth. This analysis
also provides a way to test the hypothesis of a link between the nature of dark energy and inhomogeneities. 
}


\maketitle

\section{Introduction}

A signature of hidden plane-mirror symmetry in the CMB maps has been found
in the Wilkinson Microwave Anisotropy Probe's (WMAP) temperature maps \cite{mirr1}. This is one of the non-Gaussianities among 
others reported for CMB properties (e.g. de Oliveira-Costa et al. 2004; Copi et al. 2007; 
Schwarz et al. 2004; Eriksen et al. 2004,2007; Cruz et al. 2005, 2009; Gurzadyen et al 2005,2007;
Morales \& Saez 2008).

The symmetry revealed by the inhomogeneities in the distribution of the excursion sets, 
was shown to have the highest significance at low multipoles $\ell
<5$, while the effect quickly disappears at higher multipoles \cite{mirr2}.
This effect was studied in WMAP's 94 GHz W-band maps; the foreground
reduced maps  (http://lambda.gsfc.nasa.gov/product/map/current/)
showed negligible role of Galactic contamination.

We now use another descriptor to probe the properties of the mirroring symmetry regions, 
the Kolmogorov stochasticity parameter \cite{K,Arnold,ArnoldIC} which
describes the degree of randomness of number sequences from dynamical systems or
number theory that can be applied to CMB data. As a result, a map of the degree of
randomness of CMB can be created which shows regions of high and low randomness \cite{GK_KSP,GK_K}.
Among the high randomness regions is for example the Cold Spot, southern non-Gaussian anomaly, see
\cite{Cruz}. 

Applying the Kolmogorov parameter to the mirroring symmetry regions we found that
both areas within 3$^{\circ}$ radius have coinciding values of randomness, but
that the value is different from those of other equal size regions.
This indicates either the existence of an extra randomizing factor in the
directions of the mirroring symmetry centers or different spatial depth scales in those directions.  

\section{Mirroring symmetry regions}

Mirroring symmetry has been revealed in WMAP's 3-year, and then in the 5-year 94 GHz 
W-band maps \cite{WMAP5}. This band has the highest resolution, FWHM=$0^{\circ }.21$, 
and the least contamination by the Galactic synchrotron background. 

The role of the Galactic disk was minimized via exclusion of the equatorial belt $|b|<20^{\circ }$,
and as mentioned, the foreground reduced maps were used to confirm the negligible contribution of the dust. 

The distribution of the excursion sets at the temperature interval within $|T|=45~\mu K$ 
was shown to be concentrated around almost antipodal centers (Gurzadyan et al 2007b)
\begin{eqnarray}
l&=& 94^\circ.7,\,\,\,\, b= 34^\circ.4\,\, (CE_N);  \nonumber \\
l&=& 279^\circ.8,\,\,b= -29^\circ.2\,\, (CE_S).  \nonumber
\end{eqnarray}

The locations of $CE_N$ and $CE_S$ are not special with respect the
positions of the sum of the multipoles vectors up to $\ell=8$ 
(the modulus of the vectors weighted by $1/\ell(\ell +1)$), or,  
for example, to the Cold Spot \cite{Vielva}. The positions are close
to those of the Maxwellian vectors of multipole $\ell =3$, and drift towards the equator at
higher temperature intervals.

The mirroring was studied in the context of the topology of the Universe
for a multipole dependence \cite{S93}, i.e. 
\begin{equation}
\frac{\Delta T_{\ell }(\hat{n})}{T}=\sum_{m=-\ell }^{\ell }a_{\ell m}Y_{\ell
m}(\hat{n})=\left(\frac{\Delta T}{T}\right)_{mirr}+
\left(\frac{\Delta T}{T}\right)_{non-mirr}  \label{split}
\end{equation}
where on the right hand side the first mirrored term can dominate 
at low multipoles, while the second, non-mirrored term becomes the main one at higher
multipoles. 

Partial mirroring given by Eq.(1) does not necessarily 
imply the dominance of multipoles with $|m|=\ell$, i.e. planarity,
since in the first term on
the right hand side of Eq. (\ref{split}) all $a_{\ell m}$ 
with an even value of $\ell - m$ may be non-zero.

WMAP5 maps show stronger mirroring at low multipoles, i.e. $\chi ^{2}<1.7$ for $%
\ell <5$, with monotonic weakening at higher multipoles: $\chi
^{2}>3.5$ at $\ell >10$ \cite{mirr2}. 

\section{Kolmogorov's stochasticity parameter and the symmetry regions}

Considering the CMB maps as sequences of numbers indicating the pixelized temperature,
one can represent the sky distribution of the Kolmogorov stochasticity parameter \cite{GK_KSP}.

Kolmogorov's statistic and parameter are defined \cite{K,Arnold} for $n$ independent numbers $\{X_1,X_2,\dots,X_n\}$ of  
the random variable $X$ ordered in increasing manner $X_1\le X_2\le\dots\le X_n$. 
The cumulative distribution function (CDF) of $X$ is 
$
F(x) = P\{X\le x\}, 
$
The stochasticity parameter $\lambda_n$ is defined by the limit \cite{K}
\begin{equation}
\label{KSP}
\lambda_n=\sqrt{n}\ \sup_x|F_n(x)-F(x)|\ ,
\end{equation}
where the empirical distribution function $F_n(x)$ is
\begin{eqnarray*}
F_n(x)=
0\ , & x<X_1\ ;\\
k/n\,& X_k\le x<X_{k+1},\ \ k=1,2,\dots,n-1\ ;\\
1\ , & X_n\le x\ .
\end{eqnarray*}
For any continuous CDF $F$ the convergence is uniform for
$
\lim_{n\to\infty}P\{\lambda_n\le\lambda\}=\Phi(\lambda)\ ,
$
where 
\begin{equation}
\Phi(\lambda)=\sum_{k=-\infty}^{+\infty}\ (-1)^k\ e^{-2k^2\lambda^2}\ ,\ \  \lambda>0\ ,\label{Phi}
\end{equation}
$\Phi(0)=0$, and independent of $F$ \cite{K}.

The values of the function $\Phi$ for Gaussian CDF estimated for the mirroring regions of a radius 3$^{\circ}$ centered on $CE_N$ and $CE_S$ (Figure 1)
are given in the Table 1. For comparison the values for equal size regions but situated at two other
antipodes (i.e. shifted by 180$^{\circ}$) as well as for two intermediate regions
are also represented. 

\begin{table}[ht]
\centering
\caption{Mean values for $\Phi(\lambda)$ for the mirroring (1 and 2, boldface) and 4 other regions.}
\begin{tabular}{l c c}
\hline\hline
No & Region                                     &  Mean($\Phi$) \\ [0.5ex]
\hline
1 & $CE_N$                                      & {\bf 0.7151}  \\
2 & $CE_S$                                      & {\bf 0.7155}  \\
3 & 274$^{\circ}$.7, 34$^{\circ}$.4             & 0.1877  \\
4 & 99$^{\circ}$.8,  29$^{\circ}$.2             & 0.2285  \\ 
5 & 180$^{\circ}$, 30$^{\circ}$                 & 0.1551  \\ 
6 & 180$^{\circ}$, -30$^{\circ}$                & 0.1294  \\[1ex] 
\hline
\end{tabular}
\end{table}

The situation with region 4 is indicative: its mean $\Phi =0.4397$ is decreased
to the value given in the Table when the data of a single pixel with high randomness 
(0.7) situated there are cancelled. Instead, the randomness for the mirroring regions is stable
with respect to analogical pixel elimination or shifts. Figure 2 represents the variation of $\Phi$ vs the
radius, where the difference between the mirroring regions and the others is also seen.

This shows that the regions centered at $CE_N$ and $CE_S$ possess
large scale randomness of a degree equal to each other but 3-5 times higher and
of different behavior than those for typical regions of the sky.

\begin{figure}[ht]
\centerline{\epsfig{file=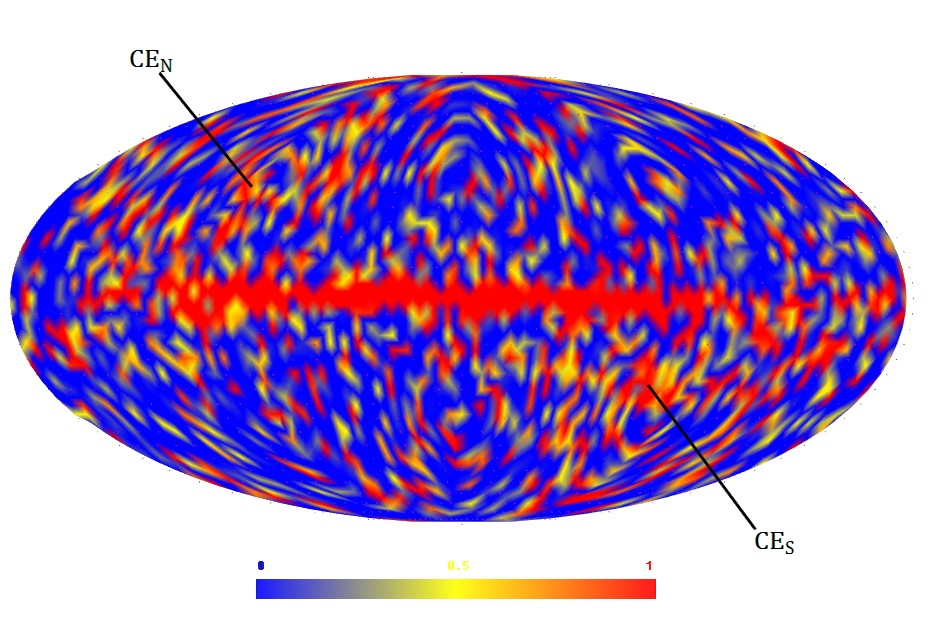,width=0.5\textwidth}} \vspace*{8pt}
\caption{Map of the degree of randomness (Kolmogorov map) for CMB WMAP's data with
indicated mirroring symmetry centers.}
\end{figure}

\begin{figure}[ht]
\centerline{\epsfig{file=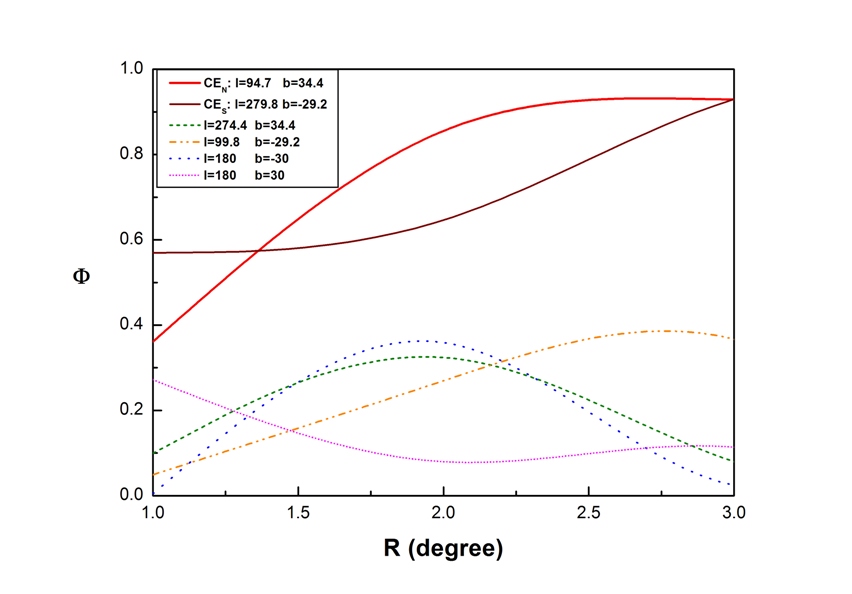,width=0.5\textwidth}} \vspace*{8pt}
\caption{The degree of randomness $\Phi$ vs the radius (in degrees) for the mirroring and 4 other areas. The difference in the
behavior in the mirroring regions (two upper curves) and the others is visible. }
\end{figure}
\section{Discussion}

Figure 1 shows the sky map of the distribution of the degree of randomness of the CMB,
as discussed above.  The centers of the mirroring symmetry are shown, which
are distinguished by higher value of Kolmogorov's parameter (randomness)
than other regions; this can be followed even in Fig.1 due to the domination of high-$\Phi$ 
structures in the vicinity of the symmetry centers.

Thus we see enhanced randomness in the mirrored regions; either
the randomizing effect is not isotropic and is dominant over those directions, or 
the line-of-sight depth is greater in those directions. Among the cosmological randomizing mechanisms can be the inhomogeneous matter distribution
in the Universe, i.e. when the voids due to their hyperbolic, diverging lensing properties can cause 
randomization of the temperature distribution in the CMB maps \cite{GK}. The voids, including
those of large scale, are among the discussed reasons also for the Cold Spot, (see Inoue \& Silk 2007; Das \& Spergel 2008).
Then, if the mirroring symmetry were of cosmological nature, it could be a result of combination of topological 
and of the integrated Sachs-Wolfe effects 
\cite{Z73,SSch75,SS75,ZS84,KS85,S93,OCSS}. 

There have been attempts to link the nature of the dark energy with light propagation effects (lensing) 
due to the inhomogeneities (see Mattsson 2007; Wiltshire 2008; Biswas \& Notari 2008; Larena et al 2008). 
Our analysis provides a way to test such a link, namely, 
if the inhomogeneities, i.e. the voids, are causing the randomization in the CMB, then the dark energy evolution 
vs redshift and the corresponding Hubble diagram have to differ in the plane-mirroring directions with respect to other directions.

Thus, with Kolmogorov's stochasticity parameter one has another indicator which 
supports the anomalous symmetry in CMB maps and acts as a quantitative 
descriptor of underlying effects. 

 {\it Acknowledgments.} We are grateful to A.A.Starobinsky for valuable discussions and the referee for helpful
comments.

\end{document}